\documentclass[twocolumn,tighten,longauthor]{myaastex62}
\shorttitle{Periodic Spectral Modulations Arise from Non-random Spacing of Spectral Absorption Lines}
\shortauthors{Michael Hippke}
\begin{document}
\title{Periodic Spectral Modulations Arise from Non-random Spacing of Spectral Absorption Lines}
\author[0000-0002-0794-6339]{Michael Hippke}
\affiliation{Sonneberg Observatory, Sternwartestr. 32, 96515 Sonneberg, Germany}
\email{michael@hippke.org}

\begin{abstract}
In recent publications, \citet{2013ApJ...774..142B,2016PASP..128k4201B,2017JApA...38...23B} claimed the discovery of ultra-short ($10^{-12}\,$s) optical pulses originating from stars and galaxies, asserted to be sent by extraterrestrial intelligence. I show that these signals are not astrophysical or instrumental in nature, but originate from the non-random spacings of spectral absorption lines. They can be shown to arise in their clearest form in synthetic solar spectra, as these do not suffer from noise.\\
\end{abstract}

\section{Introduction}
Over the last years, \citet{2010ApJ...715..589B,2010A&A...511L...6B,2012AJ....144..181B} developed and publicized a novel method, dubbed Spectral Fourier Transform (SFT), to detect and characterize ultra-short pulses from spectra. It appears that the method has gradually evolved from work by \citet{delisle1970interference,mandel1976spectral,1992ApOpt..31.3383C}. It is important to appreciate the effort to build such a new technique, a tool which may prove valuable in the future.

As a first application, SFTs were used to examine astronomical spectra from the SDSS survey of galaxies \citep{2013ApJ...774..142B}, stars \citep{2016PASP..128k4201B}, and exotic objects \citep{2017JApA...38...23B}. Astrophysical pulse detections were claimed with repetition times $\rho=1.64\times10^{-12}\,$s \citep[][stars]{2016PASP..128k4201B} and $\rho=1.09\times10^{-13}\,$s \citep[][galaxies]{2013ApJ...774..142B}. The authors speculate that the factor of $\sim 15$ between these repetition times is caused by different originators: Extraterrestrial intelligence \citep[section 7.2 in][]{2016PASP..128k4201B} versus black holes. Recent re-observations of 3 candidate stars with different equipment have failed to reproduce the signals \citep{2018arXiv181202258I}.

The \citet{BreakthroughListenReport2016} speculated that the signals may be artifacts from instrumental optics, fringing, movement of the telescope, variations in observing conditions, the process of wavelength calibration, or ``inconsistencies in the manufacture of detectors (...) of high resolution spectrographs''. This is a valid concern, as a myriad of CCD artifacts exist \citep{2016A&C....16...67D} and more are discovered regularly \citep[e.g.,][]{2018PASP..130f4504B}.

In this note, I show that the signals are caused by the non-random spacing of spectral absorption lines \citep{1924JRASC..18..373F}. They can be shown to occur in their clearest form in synthetic spectra, as these do not suffer from noise. They can also be shown to arise in high signal-to-noise ratio spectra of our sun and many other sources.

\newpage

\begin{figure*}
\includegraphics[width=\linewidth]{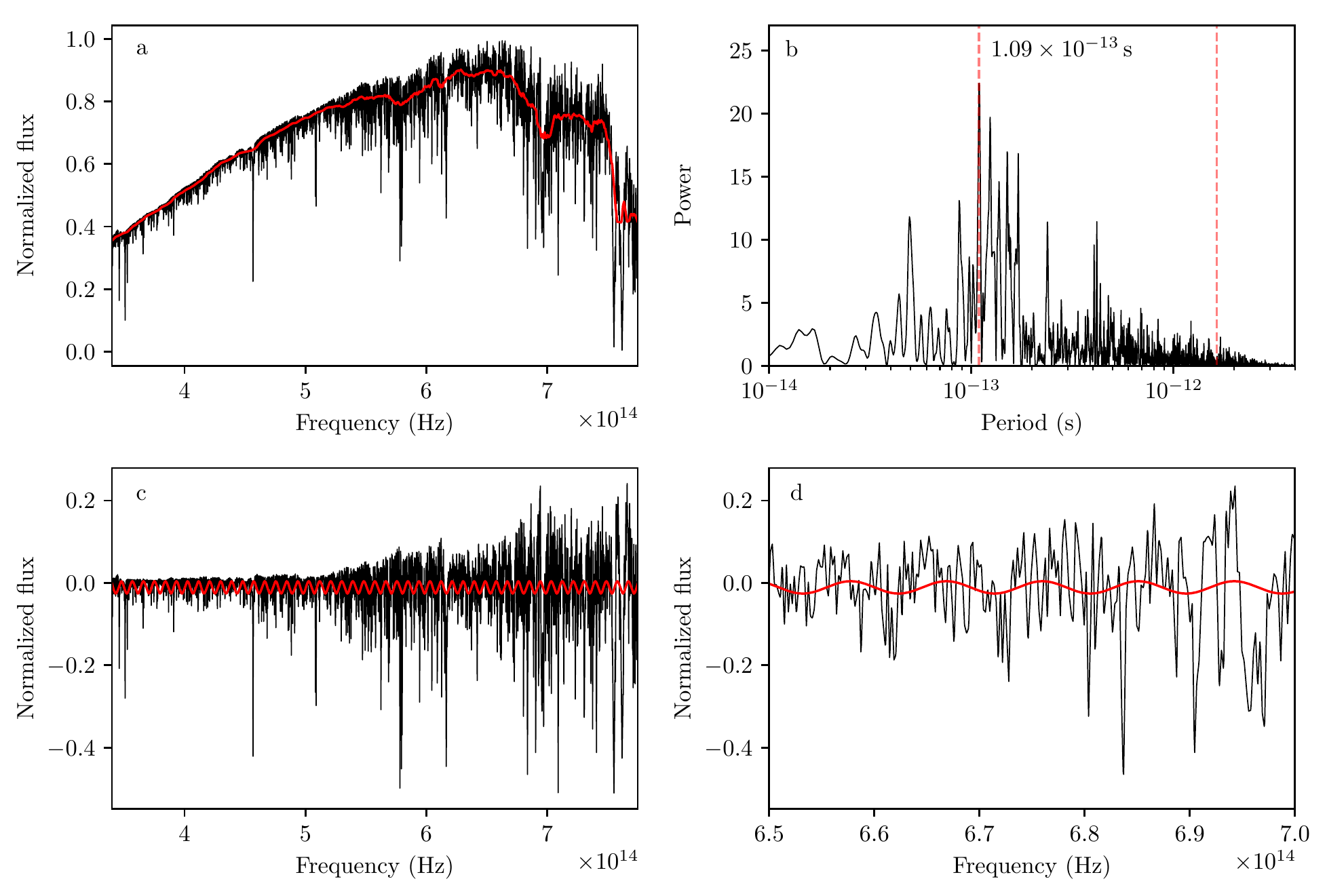}
\caption{\label{fig:kurucz}
(a): Spectral Fourier Transform of the synthetic solar spectrum by \citet{2005MSAIS...8..189K,2005MSAIS...8...73K}. A sliding median with a boxcar width of 3\,\% (red line) is fitted to the spectral flux (black line). (b): After subtracting the median from the flux, a Fourier transformation is calculated (black). The highest peak coincides with one of the claimed pulse spacings (dashed red). (c): The median-subtracted flux (black) is shown together with the best-fit sine derived from the FFT. (d): A zoom to show the fit of the sine to the data.\\}
\end{figure*}

\section{Method: Spectral Fourier transforms}
\label{sub:expl}
Most commonly, Fourier transforms are applied to time series data, to identify periodic signals. Equally, spectra can be Fourier transformed to yield periodic signals \citep{1992ApOpt..31.3383C}. A seminal summary of the method is given by \citet{2012AJ....144..181B}.

A unique feature of SFTs is their sensitivity to ultra-short periods, $\rho_{\rm min}=\lambda_{\rm min}/c\approx10^{-15}\,$s for $\lambda_{\rm min}=300\,$nm optical pulses, which is 6 orders of magnitude shorter than time-resolved ($10^{-9}\,$s) photometry using photomultipliers \citep{2010NIMPA.618..139A}. The period of the longest detectable pulses depends on $\lambda_{\rm min}$ and the spectral resolution $R$. For SDSS spectra \citep[$R=4{,}000$ at a wavelength $380<\lambda<920\,$nm,][]{2009AJ....137.4377Y,2011AJ....142...72E} we get $\rho_{\rm max}=R \, \rho_{\rm min} \sim 4\times10^{-12}\,$s.

SFTs are not sensitive to the actual pulse length $\Delta t_{\rm min}$, but to the duration between pulses ($\rho)$, so that $1/\rho$ is the pulse repetition rate. As the pulse lengths are unresolved in time, it may well be that $\Delta t_{\rm min} \ll \rho$, exacerbating the atmospheric effects of group delay dispersion. Atmospheric models indicate that turbulence smears picosecond pulses, so that it may well be impossible to use SFT-based methods on the ground. Then, the pulses could not be of astrophysical origin. Without picosecond photon counters, however, this is impossible to verify in practice \citep{2018JApA...39...74H}.

Astronomical spectra are typically sampled at equal wavelength intervals, $\lambda = c/f$ with $\Delta \lambda={\rm const}$. For a search of signals with constant temporal periods, these spectra need to be converted to equal frequency intervals $f=c/\lambda$ where $\Delta f ={\rm const}$. Afterwards, the overlaying blackbody-like flux shape needs to be subtracted out in order to perform a period search. It is adequate to use a sliding median with a boxcar window size of a few percent. Typically, sizes of 1--5\,\% give very similar results, with only a slight variation in the relative amplitudes of the peaks in the resulting Spectral Fourier transform. The SFT then produces a spectrum with power for signals with a constant temporal period, e.g., repeating laser pulses.

As the SFT method is not intuitive and requires careful implementation, I provide an open-source \texttt{Python} code\footnote{\url{http://github.com/hippke/pulses}. In the case of evenly spaced data without gaps, as is the case here, the Fourier transformation is equivalent to the Lomb-Scargle periodogram \citep{1976Ap&SS..39..447L,1981ApJS...45....1S,2018ApJS..236...16V}. The code uses the periodogram provided by Astropy \citep{2013A&A...558A..33A,2018AJ....156..123A}, but delivers identical results when swapped for SciPy's \citep{scipy} FFT implementation.} which reads spectra, performs the transformation, and creates Figure~\ref{fig:kurucz}.

\begin{figure*}
\includegraphics[width=\linewidth]{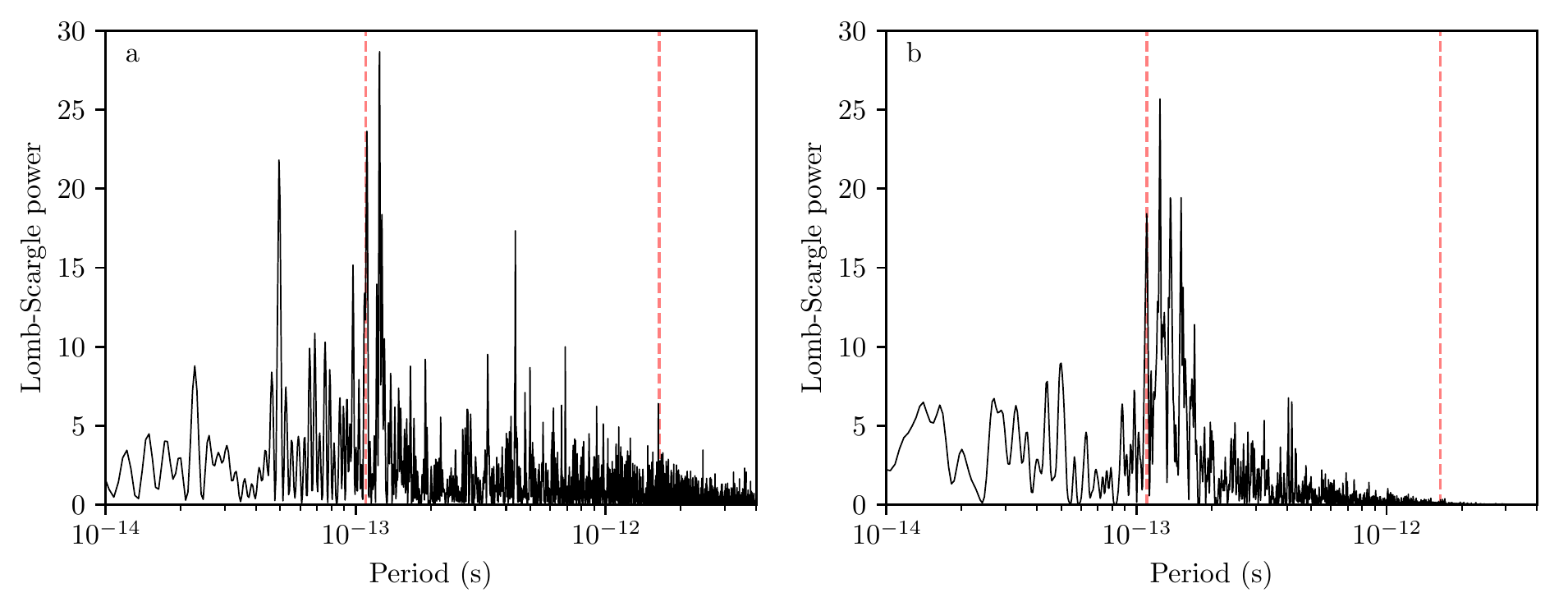}
\caption{\label{fig:stars}
(a): SFT of a solar spectrum and (b): a F9V star (TYC 2041-872-1) from SDSS found by \citet{2016PASP..128k4201B} to exhibit the signals. In both cases, a prominent peak is visible in the spectrum near one of the claimed periods.}
\end{figure*}

\section{Results: The origin of the signals}

\subsection{Synthethic spectrum}

To determine whether the signals are inherent to the data reduction method, I check the synthetic solar spectrum by \citet{2005MSAIS...8..189K,2005MSAIS...8...73K} where the telluric lines have been removed. It has the advantage that it is essentially noise-free and without instrumental issues. I down-sample and crop the resolution from $R\approx10^6$ between 300--1000\,nm to match the $R=4{,}000$ coverage between 380--920\,nm of the SDSS survey, where the signals have originally been found. Following the recipe described in the previous section, I perform the spectral Fourier transform and show the result in Figure~\ref{fig:kurucz}.

The highest spectral peak in the SFT corresponds exactly to one of the two claimed pulses at $\rho=1.09 \times 10^{-13}\,$s. The chance for this to happen by coincidence is 0.05\,\%. As there are no artificial laser pulses in a synthetic solar spectrum, it appears highly likely that such signals commonly arise as part of the processing method.

The plot of the best-fitting sine found in the period search shows that its mean is slightly below the mean of the normalized spectrum, so that the peaks of the sine fit out the flux near its nominal values, and the bottoms of the sine fit out (more or less) the absorption lines.

\subsection{Stellar spectra}
Additionally, it can be shown that the signals arise in many high signal-to-noise ratio spectra of luminous bodies with absorption lines, and quite often with the highest, or one of the highest peaks, at exactly the same $\rho=1.09 \times 10^{-13}\,$s. Figure~\ref{fig:stars} (a) shows an SFT made from a high signal-to-noise solar spectrum\footnote{Taken with the solar spectrometer installed by the University of Li\`{e}ge at the International Scientific Station of the Jungfraujoch. Available at \url{http://bass2000.obspm.fr/solar_spect.php}}. Here, the $\rho=1.09 \times 10^{-13}\,$s signal corresponds to the second highest peak when performing the same reduction as for the synthetic spectrum. The other claimed pulse spacing with $\rho=1.64\times10^{-12}\,$s coincides with a peak in the periodogram which is also significantly above the noise floor.

As a crosscheck, I have also re-reduced the SDSS spectrum of one of the stars for which \citet{2016PASP..128k4201B} found a signal (SDSS J160133.35+273355.2, Figure~\ref{fig:stars} (b)). This star was also re-observed using the optical Levy Spectrometer at the Automated Planet Finder (APF) without a detection of the signal \citep{2018arXiv181202258I}. The APF spectrum with $R \approx 100{,}000$ was not downsampled to the $R=4{,}000$ of the SDSS survey. Processing variations and different noise levels plausibly explain a non-detection of the signal.

Stars of different stellar type and metallicity exhibit different absorption lines. The power of the corresponding pulse spacings $\rho$ in the periodogram will show similar variations. Further exploration would likely not yield additional insights and may be taken as an exercise in numerology.

\subsection{Galaxy spectra}
\citet{2013ApJ...774..142B} reports a correlation of signals with redshift and argues that this excludes instrumental and data processing effects as a source. This is incorrect: the integrated flux of galaxies originates from starlight, and thus comprises the same spectral line spacings, although stacked and redshifted.

As an independent validation, I have randomly selected $12{,}000$ galactic spectra ($\sim10\,$\%) from the 6df survey \citep{2004MNRAS.355..747J}, DR3 \citep{2009MNRAS.399..683J}. Data were taken with the Anglo-Australian Observatory's UK Schmidt Telescope at a wavelength range $400<\lambda<750\,$nm at $R\sim1{,}000$, similar to SDSS, but with completely independent hardware, and placed in a different location.

Each galactic spectrum was converted to SFT using the method described in section~\ref{sub:expl} and in \citet{2012AJ....144..181B}. Redshifts were taken from \citet{2010yCat..73990683J} obtained through cross-correlation with template absorption-line spectra. SFTs were summed in redshift bins $\Delta z=10^{-4}$, typically 10--100 SFTs per bin.

\begin{figure}
\includegraphics[width=\linewidth]{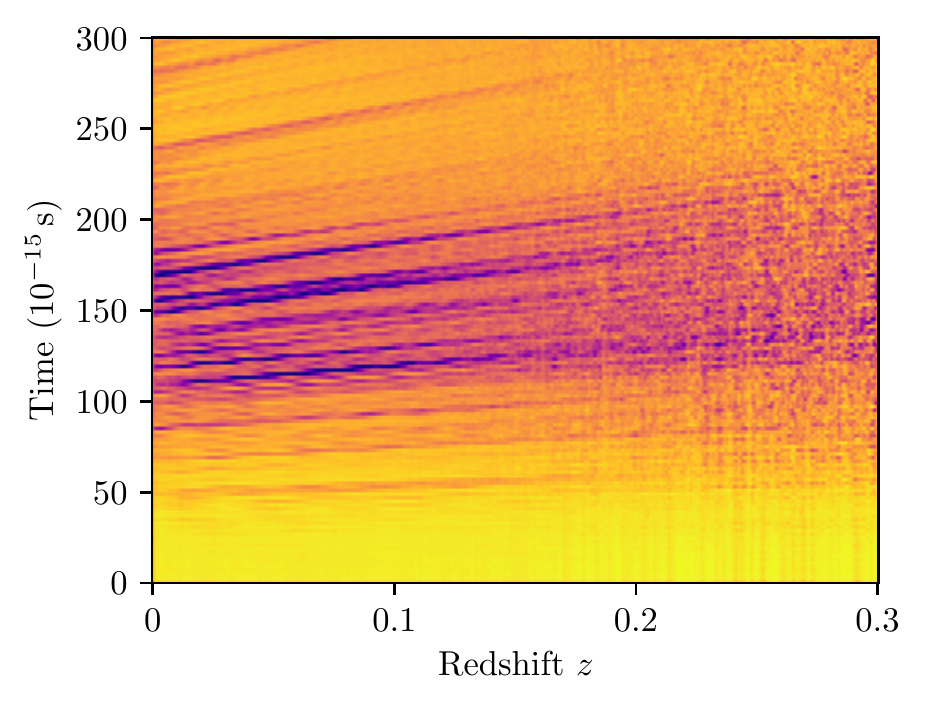}
\caption{\label{fig:fig_6df_map}
SFT stack of $12{,}000$ galaxies from the 6df survey. The stacked periodograms run from the bottom to the top in each redshift bin. The correlation of the peaks with redshift can be traced at least out to $z\sim0.2$ until noise dominates. Consequently, the peaks appear at constant periods in the rest frame of the galaxies.}
\end{figure}

The signal at $1.09\times10^{-13}\,$s is clearly visible in the resulting Figure~\ref{fig:fig_6df_map} (as $109\times10^{-15}\,$s for $z \rightarrow 0$). Additional features, undetected by \citet{2013ApJ...774..142B}, are also visible in the stack. While the vast majority of individual spectra show no significant (${\rm SNR} >6$) peaks, it is apparent that all (or a large fraction of) galaxies contribute to these patterns. It appears that, very commonly, signals are overwhelmed by noise, but can be recovered by stacking many spectra. By comparing Figures~\ref{fig:stars} and \ref{fig:fig_6df_map}, it is clear that a whole set of frequencies is detected in many stars and galaxies. This comb-like structure corresponds to a set of correlated distances of absorption lines (in units of wavelength) in the stellar and galactic spectra.

The statistical correlation of the stacked signals with redshift has been tested using linear regressions to the five strongest features for $0<z<0.2$ with a t-test. These yield positive slopes at $>6\,\sigma$ confidence in each case, for a total significance (over noise) in excess of $30\,\sigma$ confidence. This is a clear verification that the features in question are not instrumental in nature.

\section{Discussion and conclusion}
The discovery of artificial signals from outside our solar system may well require the discovery of new techniques, such as SFTs, or yet unknown tools. For this and other goals, the development of new methods will prove invaluable. My refutation of this first SFT application does not mean that the method should be abandoned; instead it should be refined. Peaks caused by the non-random spacing of absorption lines can be flagged in the future.

A natural application for the SFT method could be the template-free estimation of galactic redshift from spectra. A follow-up work could analyze the feasibility of this approach. False-positives may arise; I have found several such cases. One is SDSS J130155.84+083631.6 with a ${\rm SNR}>6$ signal near $1.09\times10^{-13}\,$s, which was classified as a galaxy with redshift $z=0.9703971$ by \citet{2010yCat..73990683J} through cross-correlation with template absorption-line spectra but is really a carbon star \citep{2013ApJ...765...12G}. The outstanding advantage of SFTs is that they do not require dedicated SETI observations, but can be performed automatically with many existing spectroscopic data originally acquired for other purposes.

The signals in question can be found in galactic spectra from the 6df survey, as well as the SDSS survey, and show a correlation with redshift. Therefore, they can not be instrumental artifacts. However, their origin is not astrophysical. They can be found in SFTs of synthetic spectra, making them a processing effect from the non-random spacings of spectral absorption lines.

\textit{Acknowledgments} I am thankful to Ermanno Borra, Brian Lacki, and Howard Isaacson for useful discussions. I thank the referee for their suggestion to examine the solar spectrum, which has lead to the results described in this manuscript.
\bibliography{references}
\end{document}